\documentclass[runningheads]{llncs}
\usepackage{cite}
\usepackage{graphicx}
\usepackage{textcomp}
\usepackage{xcolor}
\usepackage{graphicx}
\usepackage{tabularx}   

\usepackage{multirow}
\usepackage{wrapfig}
\usepackage{tablefootnote}
\usepackage{booktabs}
\usepackage{algorithm}
\usepackage{algpseudocode}
\usepackage{amsmath}
\usepackage{amssymb}

\usepackage{pdfcomment}

\def\BibTeX{{\rm B\kern-.05em{\sc i\kern-.025em b}\kern-.08em
    T\kern-.1667em\lower.7ex\hbox{E}\kern-.125emX}}

\begin{document}

\title{FLARE: Feature-based Lightweight Aggregation for Robust Evaluation of IoT Intrusion Detection}
\titlerunning{FLARE}

\author{Bradley Boswell\inst{1}\orcidID{0009-0001-1506-2533} \and
Seth Barrett\inst{1}\orcidID{0009-0007-6272-6842}\and
Swarnamugi Rajaganapathy\inst{2}\orcidID{0000-0003-4239-3677} \and
Gokila Dorai\inst{1}\orcidID{0000-0001-5825-7034}\and
Meikang Qiu\inst{1}\orcidID{0000-0002-1004-0140}}

\authorrunning{B. Boswell et al.}

\institute{Augusta University, Augusta, GA 30912, USA \email{\{brboswell,sebarrett,gdorai,mqiu\}@augusta.edu}\\
\url{https://www.augusta.edu} \and
Digital Forensics \& Artificial Intelligence Research Lab, Augusta, GA 30912, USA \email{swarnamugi@dfairlab.com}\\
\url{https://dfairlab.com} }

\maketitle

\begin{abstract}
The proliferation of Internet of Things (IoT) devices has expanded the attack surface, necessitating efficient intrusion detection systems (IDSs) for network protection. This paper presents FLARE, a feature-based lightweight aggregation for robust evaluation of IoT intrusion detection to address the challenges of securing IoT environments through feature aggregation techniques. FLARE utilizes a multilayered processing approach, incorporating session, flow, and time-based sliding-window data aggregation to analyze network behavior and capture vital features from IoT network traffic data. We perform extensive evaluations on IoT data generated from our laboratory experimental setup to assess the effectiveness of the proposed aggregation technique. To classify attacks in IoT IDS, we employ four supervised learning models and two deep learning models. We validate the performance of these models in terms of accuracy, precision, recall, and F1-score. Our results reveal that incorporating the FLARE aggregation technique as a foundational step in feature engineering, helps lay a structured representation, and enhances the performance of complex end-to-end models, making it a crucial step in IoT IDS pipeline. Our findings highlight the potential of FLARE as a valuable technique to improve performance and reduce computational costs of end-to-end IDS implementations, thereby fostering more robust IoT intrusion detection systems. 

\end{abstract}

\section{Introduction}
The IoT introduces numerous security challenges due to the vast number of interconnected devices. This poses significant risk that can compromise communication systems, disrupt critical functionalities, and exploit access control and sensitive data. According to recent guidelines published by NIST \cite{NIST}, there is a critical need for developing robust IoT security systems to address the inherent vulnerabilities in IoT devices and protect the IoT ecosystem. 
An IoT Intrusion Detection System (IoT IDS) is a security mechanism designed to monitor, analyze, detect, and classify malicious activities within an IoT ecosystem. Traditional IDS systems often use signature-based detection techniques, and their advantage is that they result in high detection rates and low false alarm rates for known attacks. However, signature-based techniques are ill-suited to the dynamic and ever-changing nature of IoT threats \cite{ids_techniques}. Modern IoT IDSs use anomaly-based detection techniques to learn normal and anomalous behaviors by analyzing network traffic and detect known and unknown attacks, making them well-suited for the ever-evolving and dynamic nature of IoT environments \cite{ids_techniques}. However, the performance of anomaly-based IDS is highly dependent on feature engineering tasks and requires designing a rich feature set that effectively characterizes network traffic patterns. Feature aggregation, as the first step in feature engineering, plays a vital role in enhancing feature selection, extraction and model performance \cite{feature_aggregation}. This can greatly optimize subsequent processes of the machine learning pipeline and contribute to higher model accuracy. In recent years, IoT IDS research has focused mainly on the feature selection and extraction process, while feature aggregation remains an active area of research \cite{research_quest}. In IoT environments, attacks often unfold over time rather than being an isolated event. The temporal context plays a vital role in these environments in detecting patterns and capturing network behaviors. While recent works have focused on addressing temporal context using deep learning models, there remains a research gap in addressing insufficient integration of meaningful time-aware features like burstiness, or flow inter-arrival patterns before model training. Temporal feature aggregation at the feature engineering stage remains a less explored area. To address aforementioned challenges, we propose FLARE, which captures session-level, flow-level and temporal network characteristics at feature engineering stage. This approach offers advantages in capturing patterns of short, sudden attack bursts effectively rather than using complex end-to-end-models that are designed to learn sequential dependencies over long time frames. Moreover, applying FLARE aggregation in feature engineering will provide a strong interpretable foundation for subsequent end-to-end learning approaches. By incorporating time-based sliding windows in our proposed feature aggregation approach, we gain deeper insights into model decision-making compared to end-to-end models, which often act as black boxes and typically require explainable AI (XAI) techniques to improve model interpretability.

Threat modeling plays a critical role in the development and evaluation of IoT intrusion detection systems, as it systematically identifies vulnerabilities, anticipates attack vectors, and formulates mitigation strategies \cite{Threat_modelling}. In this study, threat modeling is achieved by leveraging a lightweight feature aggregation methodology (FLARE) that captures session-level, flow-level and temporal network characteristics from heterogeneous IoT devices. To capture temporal patterns in the data over time, we employ a time-based sliding window \cite{sliding_window1} to capture flow-level statistics. By aggregating flow-level data and applying dimensionality reduction techniques such as PCA, the methodology highlights key features contributing to the detection of various attack types, including Denial-of-Service (DoS) and Distributed Denial-of-Service (DDoS). Through supervised learning models, this approach effectively classifies binary and multi-class attacks, demonstrating the importance of capturing fine-grained temporal dynamics and flow-based attributes in addressing IoT network intrusions.

The remainder of this paper is organized as follows: Section \ref{sec:related_work} describes the relevant background literature. Section \ref{sec:methodology} presents our proposed methodology, including feature aggregation techniques and supervised classifiers for attack classification. Section \ref{sec:experimental_eval} details the results and findings from our experimental evaluation. Section \ref{sec:conclusion} outlines the conclusions drawn from this study are presented. Finally, Section \ref{sec:future_enhancement} discusses potential avenues for future exploration and enhancement.

\section{Related Work}
\label{sec:related_work}
In this section, we review the primary factors that support the proposal of this paper. The proliferation of IoT devices in medium to large IoT ecosystems, such as home control systems and industrial IoT systems, has made these devices targets for various threats including short bursts of attacks, data theft, privacy breaches, man-in-the-middle attacks, botnet attacks, and ransomware attacks \cite{MannAttackClassifcations}. The sudden burst of attacks that is common in IoT system are referred to as a burst traffic attack, a subtype of DoS or DDoS attacks. These attacks target the inherent vulnerabilities of IoT devices, such as limited computational power and network capacity, by overwhelming them with short but intense surges of traffic or requests \cite{Lohachab}. 

A wide range of techniques and approaches have been proposed in the literature to mitigate these sudden burst of attacks in the IoT IDS. One promising approach is the integration of machine learning (ML) techniques within IoT IDS, which can enhance detection accuracy by learning from both normal and attack traffic patterns \cite{Mohammed2025}. In \cite{9031206}, the authors proposed the hybridization of the deep learning technique and the multi-objective optimization method for the detection of DDoS attacks in the IoT networks. Lightweight IoT networks are the easiest targets for attackers to introduce sudden burst of attacks. In \cite{KHANDAY2023119330}, authors highlighted a data pre-processing strategy with an ensemble feature selection algorithm to select the features by analyzing the lightweight IoT traffic patterns to detect DoS attacks.  H. Qiu and M. Qiu et al. \cite{q1,q2} proposed a new approach with machine learning \cite{A1,A2} to prevent adversarial attacks against network intrusion detection in IoT and cloud systems \cite{C1,C2,C3}. In \cite{S1}, the authors proposed contrastive learning for detecting attacks in water mark, image processing \cite{S2}, and digital twin applications \cite{B1, B2} in IoT systems. 

Current research highlights the feature aggregation process as a critical step whereby diverse low-level representations of network traffic are consolidated into richer, high-level representations. One of the primary approaches involves the aggregation of flow-based features. For instance, Adhao et.al \cite{flow_features} emphasized that flow-based aggregation not only involves the selection of informative features but also the fusion of different types of features into a unified representation. This fusion is vital for reducing noise and redundancy while emphasizing the most salient aspects of network behavior. In \cite{flow_features1},  Wu et. al. proposed that flow aggregation is achieved by grouping network packets into flows based on common characteristics such as source and destination addresses, port numbers, and temporal proximity. This aggregation results in composite flow records that encapsulate essential network behavior over defined time periods, thereby creating a more comprehensive dataset for analysis. Following flow aggregation, the authors apply latent semantic analysis, a technique developed to reveal hidden topics within textual data, to extract and quantify the underlying semantic structure present in flow records. Biyyapu et.al, \cite{flow_features2}, proposed a feature aggregation with hybrid sampling algorithm composed of ADASYN and repeated edited nearest neighbors (RENN) for sample processing to address class imbalance problem in IDS. In their article, an enhanced reptile search algorithm (IRSA) is proposed, which uses a sine cosine algorithm and Levy flight to optimally select the weight of their proposed model. Pekta\c{s} et.al \cite{flow_based_features}, proposed that grouping sequential flow records extracted from raw network traffic into two-dimensional allows deep learning architectures, specifically those combining convolutional neural networks (CNNs) with long short-term memory (LSTM) networks, to automatically learn spatial-temporal representations. 

While machine learning and deep learning models can efficiently classify IoT IDS attacks, their performance depends on well-designed feature representation and aggregation techniques. Modern intrusion detection systems designed using end-to-end models \cite{temporal1, temporal2, temporal3, temporal4, temporal5, temporal6, temporal7}, specifically for temporal feature analysis, learn feature representations dynamically from raw sequences. This increases computational complexity and requires high-end hardware like GPUs and TPUs for training. Table \ref{tab:Summary of end-to-end} summarizes the study of IDS designed using end-to-end models to process temporal features. Compared to the works outlined in this table, our work addresses these limitations by employing statistical techniques, grouping packets into flows to provide a more structured representation, aggregating network flow and temporal features before model training to reduce noise, improving interpretability by providing insights into attack patterns, and reducing computational complexities by using shallow models such as random forest, SVM, XGBoost and feedforward networks. Also, we observe from Table \ref{tab:Summary of end-to-end} that these studies have extensively focused primarily on either feature selection or feature extraction. However, they do not address incorporating feature aggregation and the potential benefits of leveraging these techniques to better detect anomalous network traffic. In our work, we demonstrate that our proposed FLARE aggregation technique serves as a foundational step in feature engineering, providing a structured representation that enhances the performance of complex end-to-end models. This makes feature aggregation a crucial component in the IoT IDS pipeline.

\begin{table}[htbp]
\begin{center}
\caption{Summary of IDS Using End-to-End Models}
\label{tab:Summary of end-to-end}
\scriptsize
\setlength{\tabcolsep}{3pt} 
\begin{tabular}{
>{\raggedright\arraybackslash}p{1cm}  
>{\raggedright\arraybackslash}p{2cm}  
>{\raggedright\arraybackslash}p{2cm}  
>{\centering\arraybackslash}p{1cm}    
>{\raggedright\arraybackslash}p{2.5cm}  
>{\raggedright\arraybackslash}p{2cm}    
}
\toprule
\textbf{Author} & \textbf{Objectives - Algorithms Used} & \textbf{Feature engineering} & \textbf{Dataset} & \textbf{Advantages} & \textbf{Limitations} \\
\midrule
Saikam et. al.(2024)\cite{temporal1} & Class Imbalance technique - Difficult set sampling \& DCGAN, Model Building: Classification - Enhanced Elman Spike NN & Feature extraction - Self Attention-based Transformer for temporal feature learning & BOT-IOT, TON-IoT, CICIDS 2019 & Addressed class imbalance, Generalization across three datasets, Self attention mechanism helps model adapt to sudden bursts of attacks & High computational cost and resource intensive, Real-time deployment challenges for low power IoT devices \\
\addlinespace
Wanjau et. al.(2024)\cite{temporal2} & Model Building: Feature extraction - bi-directional long short term memory for temporal feature learning & Feature selection - Random forest approach, Dimensionality reduction - PCA & CICIDS 2017, NSL-KDD & Proposed a hierarchical NIDS model structure that extracts discriminative and temporal characteristics from normal and malicious network traffic & Time consuming in the training phase, Increased computation cost to fuse two types of features, Difficult to process real-time data \\
\addlinespace
Lei et. al.(2021)\cite{temporal4} & Model building: CNN - convolutional layers are utilized to capture spatial correlations among different features, LSTM - to model sequential dependencies and temporal dynamics & Contribution-based feature selection & UNSW-NB15, AWID, CICIDS 2017, CICIDS 2018 & Incorporates multi-feature correlation, Integrates spatial and temporal analysis & Increased computational complexity from hybrid architecture. Demands substantial computational resources \\
\addlinespace
Kanna et. al(.2021)\cite{temporal5} & Model building: Unified model of Optimized CNN (OCNN) and Hierarchical Multi-scale LSTM (HMLSTM) & LSTM for temporal feature extraction & NSL-KDD, ISCX-IDS, UNSW-NB15 & Hyper parameter tuning using meta heuristic approach to increase the learning late & Computational overhead associated with training a unified deep learning model integrating CNN and temporal components. \\
\addlinespace
Derhab et. al.(2020)\cite{temporal7} & TCNN deep learning architecture - Convolutional Neural Network (CNN) with causal convolution & Feature space reduction and transformation using log transformation & Bot-IoT & Integrated TCNN with SMOTE-NC to handle class imbalance &  Causal convolution with larger receptive fields and additional padding require more computational resources for every forward pass  \\
\bottomrule
\end{tabular}
\end{center}
\end{table}

\section{Methodology}
\label{sec:methodology}
\begin{figure}[htbp]
    \centering
    \pdftooltip{
        \includegraphics[width=0.5\linewidth]{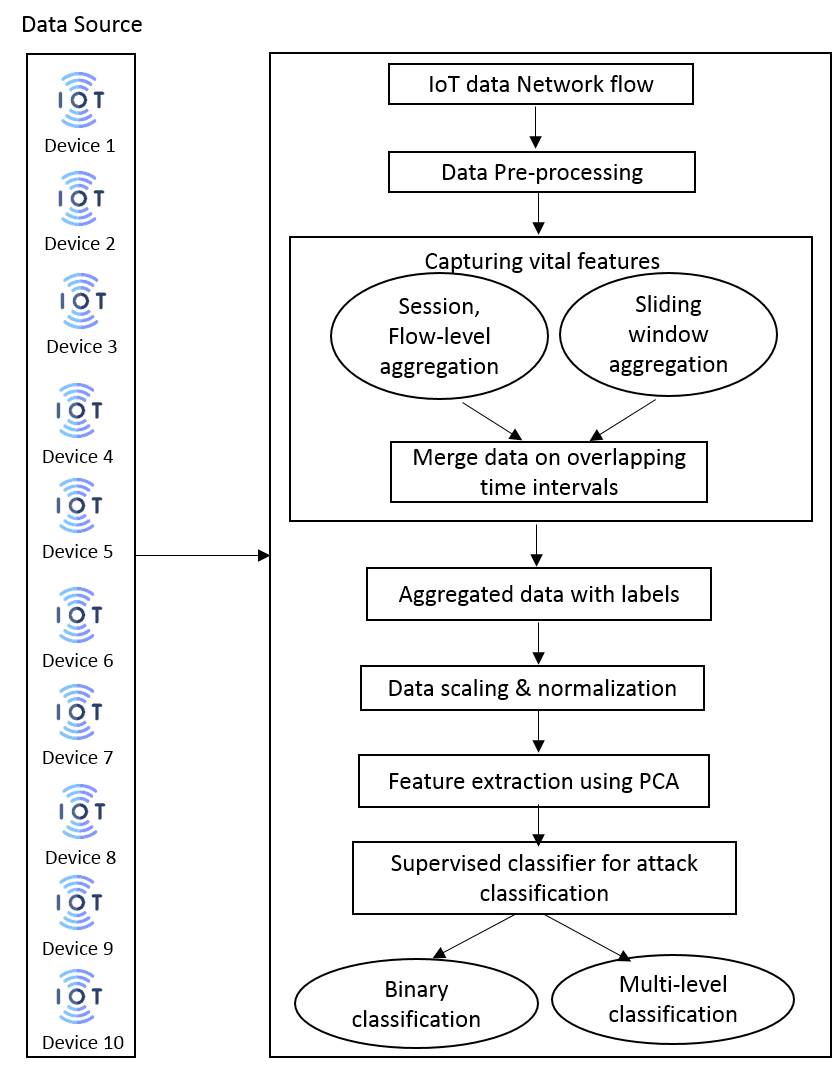}
    }{Block diagram of FLARE: illustrates preprocessing, feature aggregation (session, flow, time-based sliding windows), PCA, and supervised model classification for IoT intrusion detection.}
    \caption{FLARE - IoT Intrusion Detection System}
    \label{fig:FLARE - Block Diagram}
\end{figure}
This section outlines the primary steps for detecting and classifying attacks in our IoT intrusion detection system. We begin with data preprocessing to prepare data for the machine learning pipeline. Data collected from 10 IoT devices undergoes quality checks, cleaning, and encoding. We use unique identifiers such as timestamps to align and merge data from different systems, ensuring consistency for applying our lightweight FLARE feature aggregation technique. This aggregation produces a new dataset with highly granular temporal features, providing a fine-grained view of temporal patterns and capturing short-lived events. To extract only the most essential information, we apply Principal Component Analysis (PCA). The PCA-transformed data is then used to train supervised models for binary and multiclass attack classification. Figure \ref{fig:FLARE - Block Diagram} presents a block diagram depicting our proposed methodology for robust IoT IDS evaluation.

\subsection{Dataset Creation}
To generate a robust dataset of IoT traffic for our experiment, we set up an isolated network environment that contained a range of IoT devices. The core of the network was a TP-Link TL-WR541N router powered by OpenWrt software. Table \ref{tab:interaction_methods} lists the IoT devices chosen at this stage of the study. 
\begin{table}[htbp]
\centering
\caption{IoT devices and corresponding interaction methods}
\label{tab:interaction_methods}
\begin{tabular}{p{0.4\linewidth}p{0.5\linewidth}}
\toprule
\textbf{Device Name} & \textbf{Interaction Methods} \\
\midrule
Amazon Echo Dot, 5\textsuperscript{th} Gen. & Smartphone App \& Voice Interaction \\
Kasa Smart Wi-Fi Plug Mini & Smartphone App \& Physical Button \\
LongPlus X07 Baby Monitor & Smartphone App \\
Ring Video Doorbell, 2\textsuperscript{nd} Gen. & Smartphone App \& Device Controller \\
Google Nest Mini & Smartphone App \& Voice Interaction \\
Google Home Cam & Smartphone App \\
NiteBird Smart Bulb & Smartphone App \\
OKP K2 Robotic Vacuum & Smartphone App \\
Roborock K2 vacuum & Smartphone App \& Physical Button \\
Philips Hue Hub & Smartphone App \\
\bottomrule
\end{tabular}
\end{table}
These devices were selected to serve as examples of several operating areas, such as entertainment, security, and smart home automation. Using a Samsung Galaxy A71 5G smartphone and the corresponding companion software, each device was configured according to the manufacturer's instructions. Additionally, each device was assigned a static IP address in order to guarantee reliable packet capture that was free from issues brought on by dynamic IP changes.

Using Wireshark, we recorded benign network traffic when the IoT devices were operating normally. The specific parameters of this data collection are explained further in \cite{BarrettExploringVulnerabilities, BoswellUnravelingIoT}. These captures established a baseline for the typical communication patterns of the IoT devices. For each device in the dataset, we recorded eight-hour traffic sessions. Throughout each session, we implemented four attack methods: TCP SYN Flood, XMAS Tree Flood, UDP Flood, and HTTP Flood \cite{BarrettExploringVulnerabilities, BoswellUnravelingIoT}. Each attack was executed three times per device during the eight-hour capture period, generating a comprehensive dataset with substantial malicious traffic for analysis. All data collection sessions were conducted sequentially using Wireshark.

\subsection{Feature-based Lightweight Aggregation}
We propose a novel feature-based lightweight aggregation technique to effectively detect and classify sudden bursts of attacks in IoT intrusion detection systems. Our approach captures the highly temporal granular properties of these attack bursts by initially aggregating features for all packets within a session. This aggregation groups features based on unique network flows, which are identified by five key elements: source\_ip, destination\_ip, source\_port, destination\_port, and protocol. This aggregation allows us to summarize and consolidate traffic characteristics for each unique flow rather than each individual packet. By aggregating features for each unique session in this way, we capture both general flow-level attributes and specific direction-based attributes of forward and backward flows, such as packet length statistics, byte length statistics and the inter-arrival time between two successive packets in a flow. In order to capture the temporal dynamics of sudden bursts of attacks, we apply a sliding window to analyze packets over a specific time period, continuously sliding the time window forward in time. This allows us to collect the packets within the current window for analysis and then recalculate as new packets enter and old packets exit the window. This sliding window approach generates a series of aggregated features for each time interval, capturing characteristics such as flow rate, directional ratios, entropy metrics, and packet-level statistics. Algorithm \ref{alg:sliding_window_aggregation} shows the steps in the sliding window aggregation. We calculate shannon entropy \cite{shannon} for source\_ip and destination\_ip to measure randomness in the ip\_addresses within the window. The calculation of directional ratio features within the current window measures the forward-to-backward ratio of packets and bytes. This helps to monitor the traffic directions and identify unusual traffic behavior. Additionally, we calculate flow features within the current window to capture packet and byte transmission speed in seconds.
\begin{algorithm}[t]
\caption{Sliding Window Aggregation}
\label{alg:sliding_window_aggregation}
\begin{algorithmic}[1]
\Require \texttt{data} - the network data with indexed timestamps, 
\texttt{window\_size} - the time range of the data included in the window, \texttt{step\_size} - the stride by which the window moves forward
\Ensure \texttt{window\_list} - a data frame with aggregated features

\State Initialize \texttt{window\_list}
\State Set \texttt{start\_times} using \texttt{pd.date\_range} with \texttt{step\_size}
\For{each \texttt{start\_time} in \texttt{start\_times}}
    \State \texttt{end\_time} \(\gets\) \texttt{start\_time} + \texttt{window\_size}
    \State Extract \texttt{window} with timestamps between \texttt{start\_time} and \texttt{end\_time}
    \If{\texttt{window} is empty}
        \State \textbf{continue} to next \texttt{start\_time}
    \EndIf
    \State Calculate \texttt{flow\_rate\_features}
    \State Calculate \texttt{directional\_ratio\_features}
    \State Calculate \texttt{entropy\_features}
    \State Calculate \texttt{packet\_level\_features}

    \State Create \texttt{aggregate} dictionary with \texttt{start\_time}, \texttt{end\_time}, computed features
    \State Append \texttt{aggregate} to \texttt{window\_list}
\EndFor
\State \Return \texttt{window\_list}
\end{algorithmic}
\end{algorithm}

\begin{figure}[htbp]
    \centering
    \pdftooltip{
        \includegraphics[width=\linewidth]{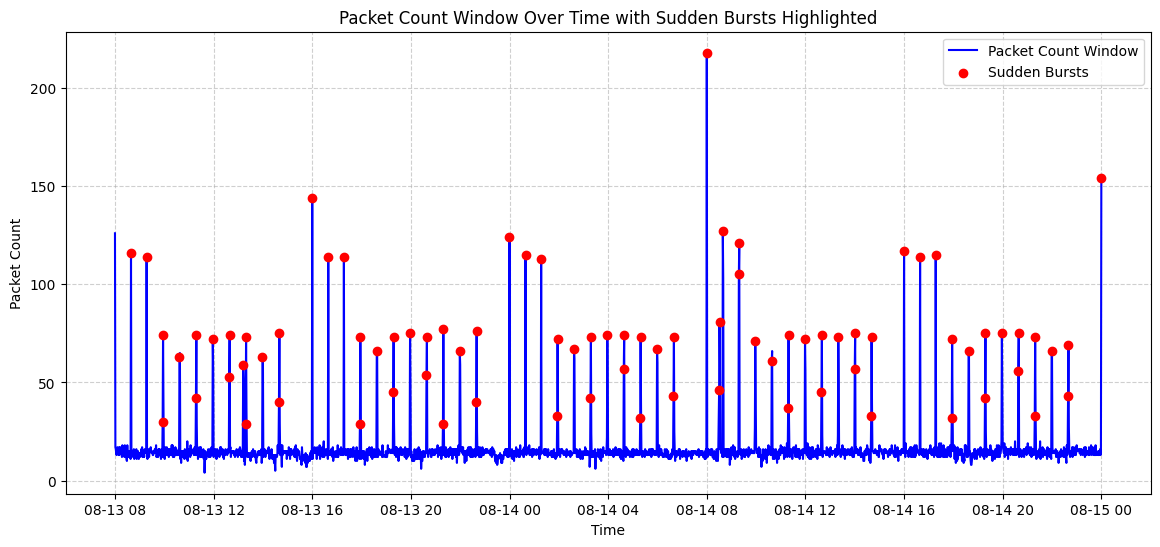}
    }{Line chart showing the packet\_count\_window attribute over time. Highlights bursts in packet counts, demonstrating temporal spikes useful for detecting IoT network anomalies.}
    \caption{Packet\_count\_window}
    \label{fig:packet_count_window}
\end{figure}

Now, having both session and sliding window aggregation, we merge these two on the overlapping time interval start\_time. For highly temporal-granularity attacks such as DoS or DDoS, aggregated features from overlapping windows share data with the previous window, providing a rolling perspective that helps detect sudden bursts of network activity. This approach offers a smoother, continuous view of network behavior, avoiding data gaps and enabling detection of attacks occurring in shorter intervals. In Figure \ref{fig:packet_count_window}, we illustrate the attribute packet\_count\_window over time with sudden bursts highlighted. For our experimental analysis, we set the step size smaller than the window size to ensure that this increases the temporal granularity and that no data are skipped.   

It is evident from our original dataset creation process that the network flows from the 10 IoT devices are originated at different time frames, and to create a unified dataset, a common attribute such as time interval is required to create an aggregated dataset from session and sliding window data that enables merging the closest matching values. In this case, we use start\_time as a common attribute, and the merge operation is performed to match start\_time in sliding window aggregation with the closest start\_time in session aggregation. By merging session and sliding window aggregations in this manner, a new aggregated dataset is produced. Algorithm \ref{alg:sliding_window_and_session_data_aggregations} shows the step in merging the sliding window and session data aggregations. This newly generated dataset is further adapted to include labels by comparing with the original dataset identified by features such as source\_ip, destination\_ip, source\_port, destination\_port and protocol. 

\begin{algorithm}[htbp]
\caption{Sliding Window and Session Data Aggregation}
\label{alg:sliding_window_and_session_data_aggregations}
\begin{algorithmic}[1]
\Require \texttt{Sliding window data frame} - aggregated data features based on a sliding time window, \texttt{session data frame} - session-based aggregated data features
\Ensure Merged aggregated dataset with enriched features 
\State Sort \texttt{sliding\_window\_data} by \texttt{start\_time}
\State Sort \texttt{session\_data} by \texttt{start\_time}
\State Initialize \texttt{aggregated\_data} as an empty list
\For{each record in \texttt{sliding\_window\_data}}
    \State Extract \texttt{start\_time} from current record
    \If{\texttt{session\_data.start\_time} $\leq$ \texttt{record.start\_time}}
        \State Merge \texttt{record} with corresponding session data
    \EndIf
    \State Append merged record to \texttt{aggregated\_data}
\EndFor
\State \textbf{return} \texttt{aggregated\_data}
\end{algorithmic}
\end{algorithm}

\subsection{Feature Extraction}
The proposed feature-based lightweight aggregation summarizes the raw network packet into high-level  features that capture key patterns and behaviors within the IoT network by combining insights from session, flow and time-based window into a set of aggregated features. We observe that some features in the aggregated data are highly correlated, and there is a necessity to minimize the feature space for robust detection and classification of attacks using supervised models. We apply dimensionality reduction through feature engineering techniques such as PCA and t-SNE. The selection of the optimal dimensionality reduction technique is determined by evaluating cluster separation quality. This is done by applying the k-means clustering algorithm \cite{10.1145/1015330.1015408} to both PCA-transformed and t-SNE-transformed data, then measuring the silhouette score to assess cluster separation. We observed that PCA outperforms t-SNE in our evaluation, with the K-means clustering on PCA-reduced data achieving a silhouette score of 0.97, compared to only 0.34 for K-means on t-SNE transformed data.
At this stage, we analyze the PCA transformed data to understand the importance of every feature and its significant contribution to capture the underlying patterns and variations in the network traffic of IoT devices. We analyze the cumulative variance, revealing that the total variance retained by the 29 components is 99.84 percent. In particular, the first three PCA components alone contribute to 50 percent of the total variance, highlighting the significant contribution to capture the underlying patterns in the data. We analyze the magnitude of feature loadings for the first three principal components to identify how features contribute to the variance across the entire dataset. These results are shown in Tables  \ref{tab:Feature_loadings_for_the_first_three_principal_components_across_the_data}  and  \ref{tab:Feature_loadings_for_the_first_three_principal_components_across_the_data2}.  Our analysis revealed that in PC1 loading the features total\_bytes\_forward, subflow\_fwd\_byts, total\_forward\_packets and fwd\_pkt\_len\_std have the highest positive loadings, and it suggests that PC1 influences the size and volume of forward traffic in terms of bytes and packets. In PC2, features like flow\_duration, fwd\_iat\_mean, bwd\_iat\_max have highest positive loadings, and this dominance reveals that time-related features play a significant role in capturing the underlying patterns. In PC3, features such as fwd\_pkt\_len\_mean, packet\_size\_mean, fwd\_pkt\_len\_min are dominant with positive loadings influenced by variations in packet size, particularly in the forward direction. 
\begin{table}[htbp]
    \centering
    \caption{Feature loadings for the first three principal components across the data}
    \begin{tabular}{lccc}
    \toprule
         \textbf{Feature}&  \textbf{PC1}&  \textbf{PC2}& \textbf{PC3}\\
         \midrule
          total\_fwd\_pkts\_window&  0.009817&  -0.008985& -0.005808\\
          total\_bwd\_pkts\_window&  0.047945&   -0.054023& 0.001519\\
         total\_fwd\_bytes\_window&  0.005998    &  -0.003896  & 0.005353\\
         total\_bwd\_bytes\_window         &  0.046971&  -0.052817& 0.002683\\
         avg\_pkt\_size\_fwd\_window             &  0.006943&  0.024710& 0.298061\\
         avg\_pkt\_size\_bwd\_window              &   0.021155&   0.009534& 0.024938\\
         flow\_duration\_window               &  0.010665&  0.000667& 0.049162\\
         packet\_count\_window            &  -0.005254&  -0.022718& -0.284901\\
         mean\_iat\_fwd\_window              &  0.011845&  0.019140 & 0.190369\\
 stddev\_iat\_fwd\_window               & 0.008918& 0.009225&0.190411\\
 
 \bottomrule
    \end{tabular}

    \label{tab:Feature_loadings_for_the_first_three_principal_components_across_the_data}
\end{table}

\begin{table}[htbp]
    \centering
    \caption{Feature loadings for the first three principal components across the data (cont.)}

    \begin{tabular}{lccc}
    \toprule
         \textbf{Feature}&\textbf{PC1}&\textbf{PC2}&\textbf{PC3}\\
         \midrule
         mean\_pkt\_length\_bwd              &  0.091830 &  0.042699& 0.091271\\
         packet\_size\_mean               &  0.078544&  0.032143& 0.318384\\
         flow\_iat\_mean              &  0.161773&  0.253869& -0.052210\\
         down\_up\_ratio            &  0.198521&   -0.163967& -0.002485\\
         subflow\_fwd\_pkts           &    0.254615&  -0.001846& -0.039997\\
         subflow\_bwd\_pkts              &  0.204008&  -0.210578& 0.007468\\
         subflow\_fwd\_byts             &  0.255975&  -0.013833& -0.032561\\
         subflow\_bwd\_byts              &  0.201297&  -0.214931& 0.008767\\
         fwd\_pkt\_len\_mean               & 0.036900& 0.046597&0.347167\\
         fwd\_pkt\_len\_max        &  0.177650     &  0.219840 &  0.090805\\
 \bottomrule
    \end{tabular}
\label{tab:Feature_loadings_for_the_first_three_principal_components_across_the_data2}
    \end{table}

From the PCA loadings, we identify the top ten features contributing to the prediction of all classes. These feature loadings are shown in Table  \ref{tab:Top_Features_Contributing_to_Prediction_of_Classes}. Our findings reveal that the features captured using the light weight aggregation technique, such as time-based forward and backward features, play a dominant role in capturing patterns and predicting our target classes. 
\begin{table}[htbp]
    \centering
    \caption{Top Features Contributing to Prediction of Classes}
    \begin{tabular}{lc}
    \toprule
           \textbf{Feature}& \textbf{Total Contribution}\\
           \midrule
           fwd\_pkt\_len\_max & 0.488295\\
           flow\_duration& 0.475808\\
           fwd\_iat\_tot& 0.475710\\
           bwd\_iat\_tot& 0.475588\\
           bwd\_iat\_mean& 0.472913\\
           fwd\_iat\_mean& 0.470734\\
           flow\_iat\_mean& 0.467851\\
           bwd\_pkt\_len\_std& 0.461786\\
           bwd\_iat\_max& 0.433369\\
  mean\_packet\_length\_forward&0.430664\\
  \bottomrule
    \end{tabular}
\label{tab:Top_Features_Contributing_to_Prediction_of_Classes}
\end{table}

\subsection{Supervised Models for Attack Classification}
Here, the objective is to train the PCA transformed data on supervised models to classify binary and multiclass attacks. After reviewing multiple studies in the literature \cite{8804727,KADRI2024101021,KAYODESAHEED20229395} and considering the characteristics of our data, our specific analysis requirements, and our focus on short-lived attack events, we identified four supervised models suitable for our classification task: Random Forest classifier, Support Vector classifier (SVC), XGBoost, and feedforward neural network. These models were selected for their ability to handle imbalanced datasets with high-dimensional features, particularly the temporal and flow-based features present in our data. The random forest classifier is best at detecting network patterns \cite{6405837} in aggregated flows and can also capture non-linear relationships that remain intact after PCA transformation. With temporal granularity, SVC is prominent in identifying fine-grained boundaries between benign and attack packets, particularly when features such as inter-packet arrival times and session duration have clear thresholds \cite{SVMnetwork}. XGboost is suitable for handling sparse, noisy data and excels at capturing temporal trends such as identifying time gaps in flows, identifying irregular bursts associated with attacks \cite{10.1117/12.3051635}. Similarly, the feedforward neural network is good at learning intricate patterns \cite{9129472} associated with flow-based and temporal dynamics effectively, making it suitable for identifying short-lived attacks. For the binary classification task, each model is trained using binary data, where each sample is labeled as either Attack or Benign, and a separate instance of each model is trained for multi-class classification task, where each sample is labeled as one of 5 attack types existing in our dataset. From the sklearn Python package, we use RandomForestClassifier and SuportVectorClassifier. The RandomForestClassifier is initialized with the default parameters. For the SVC model, we use the poly kernel parameter and enabled probability estimates. For the last two inference models, XGBoost and a feed-forward neural network, we use the xgboost and tensorflow Python packages, respectively. The feed forward neural network consists of 3 dense layers, 2 dropout layers, and uses softmax for the activation function. 
\begin{table}[htbp]
    \centering
    \caption{MSE for Flow and Temporal Features – 5s and 10s Windows}
    \begin{tabular*}{\columnwidth}{@{\extracolsep{\fill}}lcc}
        \toprule
        \textbf{Flow Features} & \textbf{5s} & \textbf{10s} \\
        \midrule
        flow\_pkts\_s & 4437.59 & 4436.67 \\
        flow\_byts\_s & 16629624.32 & 16626596.42 \\
        flow\_duration & \textbf{595660604305487.6} & 595688959677055.6 \\
        tot\_fwd\_pkts & \textbf{22363.83} & 22364.99 \\
        tot\_bwd\_pkts & \textbf{2747.70} & 2747.84 \\
        fwd\_iat\_tot & \textbf{576276543071106.2} & 576303491447727.5 \\
        bwd\_iat\_tot & 97139922598019.84 & 97138447746919.75 \\
        fwd\_pkts\_b\_avg & \textbf{0.03317} & 0.0331801 \\
        bwd\_byts\_b\_avg & \textbf{2373.65} & 2373.77 \\
        \bottomrule
    \end{tabular*}
    \label{tab:MSE_for_Flow_and_Temporal_features_5s_10s}
\end{table}
\section{Experimental Evaluation}
\label{sec:experimental_eval}
\subsection{Experiment 1: Determining Window Size }
\label{subsec:Determining window size}
For time-based sliding window aggregation, the influence of window size and step size parameters plays a vital role in capturing the temporal dynamics of the aggregated data. Proper tuning of window and step size parameters are important to determine the temporal dynamics, identify patterns, and thereby enable classifiers to classify the attacks. To determine the optimal granularity for temporal data analysis on the aggregated data, we tested four different window sizes: 5 seconds, 10 seconds, 20 seconds, and 30 seconds. To determine the optimal window size in seconds, we evaluated MSE \cite{sliding_window} for flow-based features and temporal features. MSE provides a quantitative measure for the variance of these features within a window. By comparing the MSE values across different window sizes, we assess how well the chosen window size captures patterns in the data. A low MSE indicates that the features are stable within the window, suggesting a good representation of the underlying data pattern. Tables  
\ref{tab:MSE_for_Flow_and_Temporal_features_5s_10s} and \ref{tab:MSE_for_Flow_and_Temporal_features_20s_30s}
present the MSE obtained for flow and temporal features. We selected these features for window size determination because they directly relate to packet rates, byte rates, and interarrival times, which are key indicators of bursty behavior that help detect sudden attack bursts. We observed that features like flow\_pkts\_s, flow\_byts\_s, 
bwd\_iat\_tot, yield minimum MSE for window size 30s, suggesting that it these features are suitable for detecting sustained anomalies. Features like flow\_duration, tot\_fwd\_pkts, tot\_bwd\_pkts, fwd\_iat\_tot and fwd\_pkts\_b\_avg, yield minimum MSE for window size 5s, and this suggests that these features help detect sudden bursts of attacks when set window size as 5s.

\begin{table}[htbp]
    \centering
    \caption{MSE for Flow and Temporal Features – 20s and 30s Windows}
    \begin{tabular*}{\columnwidth}{@{\extracolsep{\fill}}lcc}
        \toprule
        \textbf{Flow Features} & \textbf{20s} & \textbf{30s} \\
        \midrule
        flow\_pkts\_s & 4436.27 & \textbf{4436.11} \\
        flow\_byts\_s & 16625298.65 & \textbf{16624779.52} \\
        flow\_duration & 595701112884449.9 & 595705974319509.6 \\
        tot\_fwd\_pkts & 22365.49 & 22365.69 \\
        tot\_bwd\_pkts & 2747.90 & 2747.93 \\
        fwd\_iat\_tot & 576315041612366.9 & 576319661822777.4 \\
        bwd\_iat\_tot & 97137815620789.67 & \textbf{97137562762426.36} \\
        fwd\_pkts\_b\_avg & 0.0331808 & 0.0331811 \\
        bwd\_byts\_b\_avg & 2373.82 & 2373.84 \\
        \bottomrule
    \end{tabular*}
    \label{tab:MSE_for_Flow_and_Temporal_features_20s_30s}
\end{table}

\subsection{Experiment 2: FLARE Performance on Supervised Models}
In this section, we evaluate the performance of FLARE to classify binary and multiclass attacks in the detection of IoT IDS. Here, we have employed four supervised models to classify the attacks within the system. To ensure robustness and to evaluate generalizability of each model, we performed 10-fold cross-validation, and applied Synthetic Minority Oversampling Technique (SMOTE) on the aggregated dataset. The performance of the learning model is evaluated using metrics such as detection accuracy, precision, recall, and F1-score. For both binary and multi-classification of attacks using the supervised models, we apply window size as 5s as shown in \ref{subsec:Determining window size}, and we set step size smaller than the window size to ensure that this increases the temporal granularity and that no data are skipped. The experimental evaluation is conducted in Google Colaboratory, and we trained the four supervised models on the aggregated data with window size 5s and step size as 1s and 3s respectively. The results of the experiment are discussed separately for the binary classification and multiclass classification. 
\begin{table}[htbp]
    \centering
    \caption{Binary classification results of Random Forest and Support Vector classifiers when window\_size=5s and step\_size=1s}
    \begin{tabular}{lcccccc}
    \toprule
         &  \multicolumn{3}{c}{\textbf{Random Forest}} &  \multicolumn{3}{c}{\textbf{Support Vector}} \\
    \midrule
         &  Precision & Recall & F1 & Precision & Recall & F1 \\
    \midrule
    Benign & 1.00 & 1.00 & 1.00 & 1.00 & 1.00 & 1.00 \\
    Attack & 0.99 & 1.00 & 1.00 & 0.98 & 1.00 & 0.99 \\
    \bottomrule
    \end{tabular}
    \label{tab:bin_class_rf_svc_ws5_ss1}
\end{table}

\begin{table}[htbp]
    \centering
    \caption{Binary classification results of XGBoost and Feed Forward Neural Network when window\_size=5s and step\_size=1s}
    \begin{tabular}{lcccccc}
    \toprule
         &  \multicolumn{3}{c}{\textbf{XGBoost}} &  \multicolumn{3}{c}{\textbf{Feed Forward NN}} \\
    \midrule
         & Precision & Recall & F1 & Precision & Recall & F1 \\
    \midrule
    Benign & 1.00 & 0.99& 1.00 & 1.00 & 1.00 & 1.00 \\
    Attack & 0.86& 1.00& 0.92& 0.99 & 1.00 & 1.00 \\
    \bottomrule
    \end{tabular}
    \label{tab:bin_class_xg_ffnn_ws5_ss1}
\end{table}
\subsubsection{Binary Classification}
In Table \ref{tab:bin_class_rf_svc_ws5_ss1}, we show the binary classification performance when window\_size=5s and step\_size=1s for Random Forest and Support Vector. In Table \ref{tab:bin_class_xg_ffnn_ws5_ss1}, we show the binary classification performance where window\_size=5s and step\_size=1s for XGBoost and Feed Forward NN. 
The larger window size captures the long-term patterns and the smaller step size creates overlapping windows, increasing the number of segments for analysis. This helps us to capture short-term burst as the attack's effect will appear in multiple windows and stand out against normal traffic patterns. Therefore, with this combination, the system generated an aggregated dataset comprising a total of 23986 instances with a computational cost of nearly 23 minutes. We observe the class balance of training set as 17353 for benign and 636 for attacks. To overcome this imbalance issue and to avoid overfitting, we apply SMOTE to the training set to bring classes equal to 17353. In the test set, the number of true instances for the benign class is 5803, and for the attack class it is 194 instances. In the test set, the four trained models were able to correctly classify benign instances with an accuracy, precision, recall, and F1-score equal to 1. With respect to attack class, the random forest classifier and feed forward neural network shows effective detection of attack instance with minimal false positives and false negatives. For the other models, a small drop in precision suggests that it misclassified small instances of benign class as attacks. From this analysis, it is evident that each model performed well, without any overfitting. Indicating effective handling of class imbalances, and model generalizability on unseen data.

\begin{table}[htbp]
    \centering
    \caption{Binary classification results of Random Forest and Support Vector classifiers when window\_size=5s and step\_size=3s}
    \label{tab:bin_class_rf_svc_ws5_ss3}
    \begin{tabular}{ccccccc}
    \toprule
         & \multicolumn{3}{c}{\textbf{Random Forest}} & \multicolumn{3}{c}{\textbf{Support Vector}} \\
    \midrule
         & Precision & Recall & F1 & Precision & Recall & F1 \\
    \midrule
    Benign & 1.00 & 1.00 & 1.00 & 1.00 & 1.00 & 1.00 \\
    Attack & 0.99 & 1.00 & 0.99 & 0.99 & 1.00 & 0.99 \\
    \bottomrule
    \end{tabular}
\end{table}

\begin{table}[htbp]
    \centering
    \caption{Binary classification results of XGBoost and Feed Forward Neural Network when window\_size=5s and step\_size=3s}
    \label{tab:bin_class_xg_ffnn_ws5_ss3}
    \begin{tabular}{ccccccc}
    \toprule
         & \multicolumn{3}{c}{\textbf{XGBoost}} & \multicolumn{3}{c}{\textbf{Feed Forward NN}} \\
    \midrule
         & Precision & Recall & F1 & Precision & Recall & F1 \\
    \midrule
    Benign & 1.00 & 1.00 & 1.00 & 1.00 & 1.00 & 1.00 \\
    Attack & 0.91& 0.99& 0.95& 0.99& 1.00 & 0.99 \\
    \bottomrule
    \end{tabular}
\end{table}

Table \ref{tab:bin_class_rf_svc_ws5_ss3} shows the binary classification performance when window\_size=5s and step\_size=3s for Random Forest and Support Vector. In Table \ref{tab:bin_class_xg_ffnn_ws5_ss3}, we show the binary classification performance where window\_size=5s and step\_size=3s for XGBoost and Feed Forward NN.
In this experiment, we increased the step size by 2 compared to the previous step size 1 to analyze how this introduces changes in the overlapping windows and how it is reflected in the aggregated data instances. It is observed that it leads to less temporal representation, and reduction in the aggregated data instances as 9595, when compared to 23986 with window\_size=5s and step\_size=1s. It is also observed that it takes less computational cost around 8 minutes. In the test set, the number of true instances for the benign class is 2302, and for the attack class it is 97 instances. In the test set, all four models are able to correctly classify benign instances. With respect to the attack class, all models show a slight drop in precision and F1, and this suggests that they misclassified a small subset of the benign class as attacks. Also, the drop in precision for XGBoost, 0.91 suggests that it misclassified a small subset of instances of benign class as attacks. When comparing these results with our findings in Tables \ref{tab:bin_class_rf_svc_ws5_ss1} and \ref{tab:bin_class_xg_ffnn_ws5_ss1}, it is evident that it missed sudden bursts of attacks that occur within smaller time frames.

\begin{table}[htbp]
    \centering
    \caption{Results of Random Forest and Support Vector classifiers for multi-class classification when window\_size=5s and step\_size=1s}
    \label{tab:multi_class_rf_svc_ws5_ss1}
    \begin{tabular}{lcccccc}
    \toprule
         & \multicolumn{3}{c}{\textbf{Random Forest}} & \multicolumn{3}{c}{\textbf{Support Vector}} \\
    \midrule
         & Precision & Recall & F1 & Precision & Recall & F1 \\
    \midrule
    Benign      & 1.00 & 1.00 & 1.00 & 1.00 & 1.00 & 1.00 \\
    HTTP Flood& 0.97& 0.93& 0.95& 0.90& 0.95& 0.93\\
    TCP SYN Flood& 0.95& 0.98& 0.97& 0.96& 0.93& 0.95\\
    UDP Flood& 1.00 & 1.00 & 1.00 & 1.00 & 0.98 & 0.99 \\
    XMas Tree Flood& 0.97& 1.00 & 0.98& 1.00 & 1.00& 1.00\\
    \bottomrule
    \end{tabular}
\end{table}

\begin{table}[htbp]
    \centering
    \caption{Results of XGBoost and Feed Forward Neural Network for multi-class classification when window\_size=5s and step\_size=1s}
    \label{tab:multi_class_xgb_ffnn_ws5_ss1}
    \begin{tabular}{lcccccc}
    \toprule
         & \multicolumn{3}{c}{\textbf{XGBoost}} & \multicolumn{3}{c}{\textbf{Feed Forward NN}} \\
    \midrule
         & Precision & Recall & F1 & Precision & Recall & F1 \\
    \midrule
    Benign      
& 1.00 & 1.00 & 1.00 & 1.00 & 1.00 & 1.00 \\
    HTTP Flood
& 1.00 & 0.81 & 0.89 & 0.99& 1.00& 0.99\\
    TCP SYN Flood
& 0.90 & 1.00 & 0.95 & 1.00& 0.99& 0.99\\
    UDP Flood
& 1.00 & 1.00 & 1.00 & 1.00 & 1.00 & 1.00 \\
    XMas Tree Flood& 1.00 & 1.00 & 1.00 & 1.00 & 1.00 & 1.00 \\
    \bottomrule
    \end{tabular}
\end{table}

\subsubsection{Multi-level Classification}
For the multi-level classification evaluation, we conduct experiment using the same set of window\_size and step\_size parameters and evaluate the performance on our candidate models and, we apply SMOTE to our dataset. Tables \ref{tab:multi_class_rf_svc_ws5_ss1} and \ref{tab:multi_class_xgb_ffnn_ws5_ss1} show the performance of the multiclass classification using random forest, support vector, XGBoost and feed forward NN when window\_size=5s and step\_size=1s. Our results indicate that, on the test set, each supervised model performed well with classes such as benign, UDP Flood and XMas tree Flood. For the class HTTP Flood, the random forest, SVC, and XGBoost retain lower recall values indicating that the models missed to correctly classify this type of attack. Similarly, for the TCP SYN Flood class, the random forest model show lower precision with 0.95. This means that a small number of instances classified as TCP Flood are false positives. This misclassification occurs because attack patterns overlap between classes, making it challenging for models to distinguish them accurately. The overlapping is attributed to the larger window size, which aggregates more instances within a single window and combines different patterns. Overall, the feed forward neural network classifier demonstrates reasonable performance across all classes with the set temporal parameters compared to other models.

\begin{table}[htbp]
    \centering
    \caption{Results of Random Forest and Support Vector classifiers for multi-class classification when window\_size=5s and step\_size=3s}
    \label{tab:multi_class_rf_svc_ws5_ss3}
    \begin{tabular}{lcccccc}
    \toprule
         & \multicolumn{3}{c}{\textbf{Random Forest}} & \multicolumn{3}{c}{\textbf{Support Vector}} \\
    \midrule
         & Precision & Recall & F1 & Precision & Recall & F1 \\
    \midrule
    Benign      
& 1.00 & 1.00 & 1.00 & 1.00 & 1.00 & 1.00 \\
    HTTP Flood
& 0.67 & 0.42& 0.51& 0.53& 0.38& 0.44\\
    TCP SYN Flood
& 0.68& 0.86& 0.76& 0.64& 0.77& 0.70\\
    UDP Flood
& 1.00 & 1.00 & 1.00 & 1.00 & 0.92 & 0.96 \\
    XMas Tree Flood& 0.92& 1.00 & 0.96& 1.00 & 0.92 & 0.96 \\
    \bottomrule
    \end{tabular}
\end{table}
\begin{table}[htbp]
    \centering
    \caption{Results of XGBoost and Feed Forward Neural Network for multi-class classification when window\_size=5s and step\_size=3s}
    \label{tab:multi_class_xgb_ffnn_ws5_ss3}
    \begin{tabular}{lcccccc}
    \toprule
         & \multicolumn{3}{c}{\textbf{XGBoost}} & \multicolumn{3}{c}{\textbf{Feed Forward NN}} \\
    \midrule
         & Precision & Recall & F1 & Precision & Recall & F1 \\
    \midrule
    Benign      
& 1.00 & 1.00 & 1.00 & 1.00 & 1.00 & 1.00 \\
    HTTP Flood
& 0.96& 0.98& 0.98& 0.99& 0.98& 0.98\\
    TCP SYN Flood
& 0.98& 0.96& 0.98& 0.98& 0.99& 0.98\\
    UDP Flood
& 1.00 & 1.00 & 1.00 & 1.00 & 1.00 & 1.00 \\
    XMas Tree Flood& 1.00& 1.00 & 1.00& 1.00 & 1.00 & 1.00 \\
    \bottomrule
    \end{tabular}
\end{table}

Tables \ref{tab:multi_class_rf_svc_ws5_ss3} and \ref{tab:multi_class_xgb_ffnn_ws5_ss3}, show the performance of the multi-level classifiers random forest, support vector, XGBoost, and feed forward NN when window\_size=5s and step\_size=3s. We observe from these results that all supervised model produce less precision, recall and F1-score for the class HTTP Flood and TCP SYN Flood. This is because setting step\_size to 3s reduces the number of overlapping windows, resulting in fewer windows capturing the attack class that are short bursts. As there are minimal classes of HTTP Flood and TCP SYN Flood present, the classifiers perform well for the other classes. Overall, XGBoost and Feed forward NN yield better results for all the classes.

\begin{table}[htbp]
    \centering
    \caption{Binary Classification Results Using End-to-End Models}
    \label{tab:binary-classification-end-to-end}
    \resizebox{\textwidth}{!}{%
    \begin{tabular}{@{}llcccccc@{}}
    \toprule
    \textbf{Model} & \textbf{Class} & \textbf{Precision} & \textbf{Recall} & \textbf{F1} & \textbf{Acc.} & \textbf{Train Time} & \textbf{CPU Usage (\%)} \\
    \midrule
    \multirow{2}{*}{LSTM w{\textbackslash}o FLARE} 
        & Benign & 0.99 & 0.98 & 0.98 & \multirow{2}{*}{0.97} & \multirow{2}{*}{1449.76 s} & Before: 34.8 \\
        & Attack & 0.79 & 0.85 & 0.82 & & & After: 65.4 \\
    \addlinespace
    \multirow{2}{*}{LSTM with FLARE} 
        & Benign & 1.00 & 1.00 & 1.00 & \multirow{2}{*}{0.99} & \multirow{2}{*}{59.33 s} & Before: 33.5 \\
        & Attack & 0.99& 1.00 & 1.00& & & After: 48.9 \\
    \addlinespace
    \multirow{2}{*}{BI-LSTM w{\textbackslash}o FLARE} 
        & Benign & 0.99 & 0.98 & 0.98 & \multirow{2}{*}{0.97} & \multirow{2}{*}{1603.38 s} & Before: 38.7 \\
        & Attack & 0.78 & 0.87 & 0.83 & & & After: 74.4 \\
    \addlinespace
    \multirow{2}{*}{BI-LSTM with FLARE} 
        & Benign & 1.00 & 1.00 & 1.00 & \multirow{2}{*}{0.99} & \multirow{2}{*}{59.27 s} & Before: 14.7 \\
        & Attack & 0.99 & 1.00 & 1.00 & & & After: 42.8 \\
    \bottomrule
    \end{tabular}
    }
\end{table}
\begin{table}[htbp]
    \centering
    \caption{Multi-Classification Results of LSTM with and without FLARE}
    \label{tab:multi_class_LSTM_}
    \begin{tabular*}{\columnwidth}{@{\extracolsep{\fill}}lcccccc}
    \toprule
         & \multicolumn{3}{c}{\textbf{LSTM w{\textbackslash}o FLARE}} & \multicolumn{3}{c}{\textbf{LSTM with FLARE}} \\
    \midrule
         Classes& Precision & Recall & F1 & Precision & Recall & F1 \\
    \midrule
    Benign      & 0.98& 0.98& 0.98& 1.00& 1.00& 1.00\\
    HTTP Flood& 0.52& 0.40& 0.45& 0.98& 0.98& 1.00\\
    TCP SYN Flood& 0.57& 0.67& 0.62& 1.00& 1.00& 1.00\\
    UDP Flood& 0.77& 0.77& 0.77& 1.00& 1.00& 1.00\\
    XMas Tree Flood& 0.93& 0.97& 0.95& 1.00& 1.00& 1.00\\
    \bottomrule
    \end{tabular*}
\end{table}
\begin{table}[htbp]
    \centering
    \caption{Multi-Classification Results of BI-LSTM with and without FLARE}
    \label{tab:multi_class_BiLSTM}
    \begin{tabular*}{\columnwidth}{@{\extracolsep{\fill}}lcccccc}
    \toprule
         & \multicolumn{3}{c}{\textbf{BI-LSTM w{\textbackslash}o FLARE}} & \multicolumn{3}{c}{\textbf{BI-LSTM with FLARE}} \\
    \midrule
         Classes& Precision & Recall & F1 & Precision & Recall & F1 \\
    \midrule
    Benign      & 0.98& 0.98& 0.98& 1.00& 1.00& 1.00\\
    HTTP Flood& 0.57& 0.28& 0.37& 1.00& 0.99& 0.99\\
    TCP SYN Flood& 0.54& 0.76& 0.63& 0.99&1.00& 0.99\\
    UDP Flood& 0.74& 0.82& 0.78& 1.00& 1.00& 1.00
\\
    XMas Tree Flood& 0.94& 0.97& 0.95& 1.00& 1.00& 1.00\\
    \bottomrule
    \end{tabular*}
\end{table}
\begin{table}[htbp]
    \centering
    \caption{End-to-End Models: Computational resource utilization for multi-class classification.}
    \label{tab:End-to-End Model computational resource utilization for Multi-class classification}
    \begin{tabular*}{\columnwidth}{@{\extracolsep{\fill}}lccc@{}}
        \toprule
        \textbf{End-to-End Model} & \textbf{Acc.} & \textbf{Train Time (s)} & \textbf{CPU Usage (\%)} \\
        \midrule
        \multirow{2}{*}{LSTM w{\textbackslash}o FLARE} 
            & \multirow{2}{*}{0.9597} & \multirow{2}{*}{866.23} & Before: 15.7 \\
            & & & After: 65.8 \\
        \addlinespace
        \multirow{2}{*}{LSTM with FLARE} 
            & \multirow{2}{*}{0.9992} & \multirow{2}{*}{77.44} & Before: 28.2 \\
            & & & After: 21.4 \\
        \addlinespace
        \multirow{2}{*}{Bi-LSTM w{\textbackslash}o FLARE} 
            & \multirow{2}{*}{0.9588} & \multirow{2}{*}{686.63} & Before: 42.5 \\
            & & & After: 65.6 \\
        \addlinespace
        \multirow{2}{*}{Bi-LSTM with FLARE} 
            & \multirow{2}{*}{0.9969} & \multirow{2}{*}{102.12} & Before: 26.0 \\
            & & & After: 67.3 \\
        \bottomrule
    \end{tabular*}
\end{table}
\subsection{Experiment 3: FLARE Performance on End-to-End Models}
We conducted an experiment to demonstrate how our proposed FLARE aggregation technique improves model performance. This experiment compared algorithms from related work, such as LSTM and BI-LSTM, both with and without the FLARE aggregation technique. The results support our claim that applying FLARE as an initial step in the feature engineering process, before feature selection and extraction, creates a more structured data representation and enhances the performance of complex end-to-end models. For binary classification, we designed an LSTM and bidirectional LSTM model with 64 units as layer one to capture temporal dependencies, and second layer with 32 units to process the sequential data. We set the dropout as 0.2 to reduce overfitting, dense layer 16 units with RELU activation, followed by a softmax output layer for binary classification. In Table \ref{tab:binary-classification-end-to-end}, LSTM and Bi-LSTM without FLARE takes a preprocessed input instance of 125899 to perform binary classification. In LSTM and Bi-LSTM with FLARE, we incorporated the proposed aggregation as an initial feature engineering step, thus aggregated the instances to 23986 by setting window size to 5s and step size as 1s, and used SMOTE for balancing the classes. Our findings show that both algorithms, when augmented by our FLARE aggregation technique, yielded better performance and also used less computational resources compared to LSTM and BiLSTM without the FLARE aggregation technique. 

For multi-classification, we designed the LSTM and BI-LSTM architectures with a first layer of 128 units to process the input sequence and learn temporal patterns. A dense layer with 64 units functions as a fully connected layer with ReLU activation for feature learning. Two dropout layers, each set to 0.3, are used to reduce overfitting and enhance generalization. The network concludes with a dense layer using softmax activation for multi-class classification. The performance analysis of LSTM and BI-LSTM for multi-classification is shown in Tables \ref{tab:multi_class_LSTM_} and \ref{tab:multi_class_BiLSTM}, respectively. It is observed that the performance of end-to-end models with FLARE aggregation produced higher accuracy, and a relatively low amount of false positives or false negatives compared to end-to-end models without FLARE aggregation. Additionally, we analyzed the computational utilization of both end-to-end models. The results of this analysis are shown in Table \ref{tab:End-to-End Model computational resource utilization for Multi-class classification}. These results indicate that the training time for models with FLARE are significantly reduced when compared to models that operated without FLARE aggregation. This observation underscores a critical advantage in adopting FLARE as an aggregation technique in the context of IoT IDS.

\section{Conclusion}
\label{sec:conclusion}
The growing prevalence of IoT devices has significantly increased the risk of security attacks, particularly sudden bursts of attacks like DoS, necessitating efficient and timely intrusion detection systems. While the literature has introduced and analyzed numerous feature engineering approaches, including feature selection and extraction techniques, limited attention has been given to feature aggregation as a potential method to enhance anomaly detection in IoT systems. In this work, we proposed FLARE, a novel feature aggregation technique that captures vital features from the session, flow, and temporal dynamics of IoT intrusion detection datasets. To determine the optimal window size for time-based sliding windows, we analyzed and evaluated MSE for flow and temporal features and applied PCA for feature extraction. We addressed class imbalance by using SMOTE on the training set. We trained four supervised models and two deep learning models to perform both binary attack detection and multi-class attack classification. Our experimental evaluation revealed that a step size smaller than the window size, particularly window\_size=5s and step\_size=1s, increased the temporal granularity and enhanced the robustness of FLARE. Our analysis demonstrated that end-to-end models incorporating FLARE aggregation showed better performance and reduced computational complexity compared to models operating without FLARE. 

\section{Future Enhancement}
\label{sec:future_enhancement}
While our research has demonstrated the importance of feature aggregation techniques in preserving vital features for classifying sudden bursts of attacks in IoT intrusion detection systems, several avenues remain for future exploration and enhancement. First, to further improve the effectiveness of the aggregated dataset and model performance, advanced aggregation and class balancing techniques could be explored. These include class-preserving aggregation methods such as stratified or adaptive window-based aggregation to ensure all classes, especially minority ones, are well-represented. Second, post-aggregation balancing techniques like synthetic sampling or resampling methods could address any residual imbalance. There is also opportunity to explore replacing fixed parameters with methods that dynamically set temporal parameter values suitable for more dynamic IoT networks. Lastly, this work opens possibilities for incorporating additional machine learning and deep learning models that can enhance the system's ability to detect and classify attacks effectively.

%
%
%
\bibliographystyle{splncs04}
\bibliography{main}
%
\end{document}